\documentclass[sn-mathphys]{sn-jnl}% Math and Physical Sciences Reference Style
\jyear{2021}%

\theoremstyle{thmstyleone}%
\usepackage{graphicx}% Include figure files
\usepackage{bm}% bold math
\usepackage{commath}

\usepackage{subcaption}
\theoremstyle{thmstyletwo}%

\theoremstyle{thmstylethree}%

\newcommand{\be}{\begin{equation}}
\newcommand{\ee}{\end{equation}}
\newcommand{\bea}{\begin{eqnarray}}
\newcommand{\eea}{\end{eqnarray}}
\newcommand{\lb}{\label}

\newcommand{\bv}{{\bf v}}
\newcommand{\bu}{{\bf u}}
\newcommand{\bw}{{\bf w}}
\newcommand{\bk}{{\bf k}}

\newcommand{\bx}{{\bf x}}

\newcommand{\bG}{{\bf G}}

\newcommand{\bJ}{{\bf J}}

\newcommand{\bD}{{\bf D}}

\newcommand{\bR}{{\bf R}}

\newcommand{\bphi}{{\mbox{\boldmath $\phi$}}}

\newcommand{\boeta}{{\mbox{\boldmath $\eta$}}}

\newcommand{\bsigma}{{\mbox{\boldmath $\sigma$}}}

\newcommand{\grad}{{\mbox{\boldmath $\nabla$}}}
\newcommand{\bdot}{{\mbox{\boldmath $\cdot$}}}

\newcommand{\bzed}{{\mbox{\boldmath $0$}}}

\newcommand{\baro}{\bar{{\rm o}}}

\DeclareMathOperator\erf{erf}
\begin{document}

\title{The Kraichnan Model and Non-Equilibrium Statistical Physics of Diffusive Mixing}

%% NatureName	->                \dgr{MSc, PhD}}\email{iauthor@gmail.com}
%%=============================================================%%

\author*[1,2]{\fnm{Gregory} \sur{Eyink}}\email{eyink@jhu.edu}

\author[1]{\fnm{Amir} \sur{Jafari}}\email{elenceq@jhu.edu}
\equalcont{These authors contributed equally to this work.}

\affil*[1]{\orgdiv{Department of Applied Mathematics and Statistics}, \orgname{The Johns Hopkins University}, \orgaddress{\street{3400 N. Charles St.}, \city{Baltimore}, \postcode{21218}, \state{Maryland}, \country{USA}}}

\affil[2]{\orgdiv{Department of Physics and Astronomy}, \orgname{The Johns Hopkins University}, \orgaddress{\street{3400 N. Charles St.}, \city{Baltimore}, \postcode{21218}, \state{MD}, \country{USA}}}


\abstract{ We discuss application of methods from the Kraichnan model of turbulent advection to the study 
 of non-equilibrium concentration fluctuations arising during diffusion in liquid mixtures at 
 high Schmidt numbers. This approach treats nonlinear advection of concentration fluctuations 
 exactly, without linearization. Remarkably, we find that static and dynamic structure functions
 obtained by this method reproduce precisely the predictions of linearized fluctuating hydrodynamics. 
 It is argued that this agreement is an analogue of anomaly non-renormalization which does not,
 however, protect higher-order multi-point correlations. The latter should thus yield
 non-vanishing cumulants, unlike those for the Gaussian concentration fluctuations predicted by linearized theory. }


\maketitle

\section{\label{Intro}Introduction}

The Kraichnan model \cite{kraichnan1968small,kraichnan1974convection,Kraichnan94} of a passive scalar 
advected by a turbulent fluid has achieved a paradigmatic status in the modern 
theory of turbulence \cite{falkovich2001particles}. A breakthrough in the
analytical calculation of anomalous scaling for scalar structure functions was 
initiated by the work of Gaw\c{e}dzki and Kupiainen \cite{gawedzki1995anomalous},
who exploited an expansion in the smoothness exponent $\xi$ of the velocity field 
around the rough limit $\xi=0.$ This was followed very closely by alternative approaches based on 
expanding in powers of the inverse $1/d$ of space dimension \cite{chertkov1995normal} and 
in powers of $2-\xi$ for $\xi$ near $2$ \cite{shraiman1995anomalous}. The Kraichnan model 
has been the source of fundamental new concepts in turbulence theory such as Lagrangian 
slow modes and spontaneous stochasticity \cite{bernard1998slow} and it has proved 
also a valuable testing ground for general renormalization group methods \cite{gawedzki1997inverse,antonov2006renormalization,kupiainen2007scaling}
The contemporaneous reviews of Gaw\c{e}dzki \cite{Gawedzki1997,gawedzki2002easy,gawedzki2002soluble,cardy2008non}
written as progress reports still provide some of the most insightful and lucid expositions of this body of work. 

More recently, the Kraichnan model has found an entirely new arena of application 
in the non-equilibrium statistical mechanics of diffusive mixing. The paper of 
Donev, Fai and vanden-Eijnden \cite{donev2014reversible} (hereafter DFV) has shown 
that the equations for the scalar concentration field in a binary fluid mixture,  
which describe its advection by thermal fluctuations of the velocity in an otherwise 
quiescent fluid, reduce in the high-Schmidt limit to a version of the Kraichnan model
and yield an economical scheme for efficient numerical computation. The subsequent
work \cite{eyink2022high} applied the DFV theory analytically to investigate the
effects of thermal noise on high-Schmidt turbulent mixing and showed, in the 
process, that the many powerful mathematical methods devised to treat turbulent advection 
in the Kraichnan model can be applied also to mixing by thermal fluctuations.  
In the present paper we shall explain further this application and present a 
few initial results of our ongoing investigation. This subject 
seems very suitable for the special journal issue in memory of Krzysztof Gaw\c{e}dzki. 
Among his very wide-ranging research pursuits, Gaw\c{e}dzki himself made important contributions 
to mathematical physics of non-equilibrium statistical mechanics for diffusion processes 
\cite{chetrite2008Afluctuation,chetrite2008Bfluctuation,chetrite2009eulerian,gomez2009experimental}. 
We therefore offer this work in tribute to him and to express our deep gratitude
for his scientific leadership and personal friendship. 

Although diffusive mixing in a laminar fluid may appear to have little in common 
with turbulence, the physical analogies become apparent on closer inspection. In both problems,
for example, there are scale-invariant fluctuations leading to correlations with power-law decay. 
Concentration correlations for steady-state diffusion with a mean concentration gradient imposed 
by the Soret effect were first studied by Law and Nieuwoudt \cite{law1989noncritical,nieuwoudt1990theory}, 
who predicted long-ranged correlations characterized by a power-law divergence 
$\propto k^{-4}$ of the static structure function at low wave-numbers. This is a special 
case of the generic long-range fluctuations expected in non-equilibrium statistical steady-states 
\cite{dorfman1994generic,grinstein1995generic}. It was subsequently found by Vailati, Giglio, and 
co-workers that similar correlations occur also in isothermal free diffusion of a blob of concentration, 
where the mean concentration gradient is co-evolving with the fluctuations 
\cite{vailati1997giant,brogioli2000universal,croccolo2007nondiffusive}. 
These non-equilibrium concentration fluctuations have been termed ``giant'' since they are 
orders of magnitude larger than thermal equilibrium fluctuations \cite{vailati1997giant}.
The scale of the fluctuations is also macroscopic, cut off only by bouyancy effects
\cite{segre1993nonequilibrium,vailati1997giant,brogioli2000universal,croccolo2007nondiffusive}, 
and, in low gravity environments, quenched only by the finite size of the sample 
\cite{ortiz2004nonequilibrium,vailati2011fractal,Cerbino2015,croccolo2016shadowgraph}. 
Indeed, as in turbulent eddy transport, diffusive flux in liquids can be understood 
to be generated entirely by the non-equilibrium fluctuations
\cite{brogioli2000diffusive,donev2011diffusive,donev2011enhancement,donev2014reversible}. 
Thus, the diffusion coefficient measured in a macroscopic experiment is in fact 
an ``eddy diffusivity'' due to advection of concentration by thermal velocity fluctuations. 

The standard equations to describe this physics are the Landau-Lifschitz fluctuating 
hydrodynamics equations for a binary mixture  in a solvent at rest
\cite{Cerbino2015,brogioli2016correlations}: 
\bea\label{Momentum100}
\rho\partial_t {\bf v}'&=&-\grad p'+ \eta \triangle {\bf v}'+\grad\bdot \Big(\sqrt{ 2\eta k_B T}\; {\boldsymbol\eta}({\bf x}, t) \Big),
\eea 
\be\label{Passive110}
\partial_t c=-\bv'\bdot\grad c +\grad\bdot \left(D \grad c\right).
\ee
The fluctuation velocity $\bv'$ is assumed incompressible, $\grad\bdot\bv'=0,$ for constant mass 
density $\rho$ of the fluid, with this constraint enforced in \eqref{Momentum100} by the pressure $p'.$  
The tensor field ${\boldsymbol \eta}({\bf x}, t)$ 
is Gaussian white-noise, symmetric and traceless, representing thermal fluctuating stress, with zero mean and covariance 
\begin{eqnarray}\lb{etav} 
\langle  \eta_{ij}({\bf x}, t) \eta_{kl}({\bf x'}, t')\rangle&=&(\delta_{ik}\delta_{jl}+\delta_{il}\delta_{jk}-{2\over 3} \delta_{ij}\delta_{kl})\delta^3({\bf x-x'})\delta(t-t').
\end{eqnarray}
The prefactor $\sqrt{ 2\eta k_B T}$ involving the shear viscosity $\eta$ and temperature $T$ is dictated by the 
fluctuation-dissipation relation so that the correct Gibbs equilibrium distribution is obtained 
for the equal-time velocity statistics with energy equipartition among wave-number modes 
(e.g., see Appendix A in \cite{eyink2021dissipation}). We have neglected here, as in
\cite{Cerbino2015,brogioli2016correlations},  the similar thermal noise term in the equation 
\eqref{Passive110} for the concentration $c,$ since the equilibrium fluctuations which it 
generates are far smaller than the giant concentration fluctuations (GCF's) of primary interest. 
Note that $D$ in \eqref{Passive110} is the molecular diffusivity and that we could 
also add to the diffusive mass current $\bJ_D=-D\grad c$ a term $\bJ_T=-D c(1-c)S_T \grad T$
corresponding to non-equilibrium mass flux driven by a temperature gradient $\grad T,$
with $S_T$ the Soret coefficient \cite{degroot2013non}. 

The present standard theory \cite{vailati1998nonequilibrium,brogioli2000diffusive,brogioli2016correlations}
based upon {\it linearized fluctuating hydrodynamics} proceeds by writing 
an equation for the mean concentration $\bar{c}$:
\be\label{PassiveMean}
\partial_t \bar{c}= \grad\bdot \left(D \grad \bar{c} \right),
\ee
by then defining the concentration fluctuation $c'=c-\bar{c},$ and by finally linearizing \eqref{Passive110} 
to obtain 
\be\label{PassiveFluc}
\partial_t c'=-\bv'\bdot\grad \bar{c} +\grad\bdot \left(D \grad c'\right),
\ee
with the term $-\bv'\bdot \grad c'$ which is quadratic in fluctuations discarded. The latter linearized
equation may then be solved exactly in the case of a stationary, homogeneous mean concentration gradient 
$\grad\bar{c}$ \cite{brogioli2000diffusive,brogioli2016correlations} or else approximately when the solution 
$\bar{c}(\bx,t)$ of \eqref{PassiveMean} has a gradient $\grad \bar{c}(\bx,t)$ which is dependent upon $\bx$ and $t$ 
\cite{vailati1998nonequilibrium}. This simple approach has yielded quite accurate predictions 
of experimental observations for free diffusion processes, except for minor deviations at early times when concentration gradients are relatively large \cite{croccolo2007nondiffusive}. However, current experimental efforts in low-gravity environments such as the NEUF-DIX space project \cite{baaske2016neuf,vailati2020giant} 
are now exploring problems with large concentration gradients, high concentrations, and transient processes 
where the basic assumptions of the standard theory are invalid. 

The exact asymptotic theory of DFV  \cite{donev2014reversible} is here very attractive because 
it is fully nonlinear and the basic assumption of high Schmidt number is well-satisfied 
in most of the liquid mixtures experimentally considered. An obstacle to serious consideration  
of this approach may have been the puzzling numerical result obtained in \cite{donev2014reversible}
for the single-time structure function $S(\bk,t)$ in a free-diffusion experiment. Rather than 
the expected scaling $S(\bk,t)\propto k^{-4},$ the numerical implementation of the DFV theory 
in \cite{donev2014reversible} produced a result apparently more consistent with $S(\bk,t)\propto k^{-3}.$
On the other hand, it has now been shown by an exact mathematical analysis \cite{eyink2022high}
that the DFV theory does yield $S(\bk)\propto k^{-4},$ in close agreement with linearized theory,  
at least for a statistical steady state with random injection of concentration fluctuations. It was 
speculated in \cite{eyink2022high} that the simulation by DFV in \cite{donev2014reversible} was
not run for a sufficient time to observe the correct $S(\bk,t)\propto k^{-4}$ scaling or perhaps 
had an insufficient wavenumber range to clearly identify the power law exponent. Thus, the 
DFV theory does appear to give results consistent with linearized fluctuating hydrodynamics 
and, more importantly, in agreement with laboratory experiment. 

In the present paper we further support this conclusion of \cite{eyink2022high} by solving 
exactly for the static and dynamic structure functions of the DFV theory in the situation 
of greatest experimental interest, where the GCF's are generated by an imposed concentration 
gradient interacting with thermal velocity fluctuations. We first briefly review the DFV 
theory and then derive useful forms of the closed equations for the mean concentration and 
for the 2-point correlation function. We next solve these equations analytically for the simplest 
case of a stationary, homogeneous mean concentration gradient. The somewhat surprising 
result we obtain is that the nonlinear DFV theory yields {\it precisely} the same result 
for the structure function as does the standard linearized theory 
\cite{vailati1998nonequilibrium,brogioli2000diffusive,brogioli2016correlations}.
As we shall discuss, this result may be considered as a simple example of an ``anomaly non-renormalization 
theorem'', which helps to explain the remarkable success of linearized fluctuating hydrodynamics 
in describing the observations from experiment and numerical simulation for 2nd-order statistics
in high-Schmidt liquid mixtures. However, higher-order correlation functions of the concentration fluctuations are not 
protected by such a result and are unlikely to be given correctly by the Isserlis-Wick
theorem for a Gaussian random field in terms of the 2nd-order order correlation function. 
We discuss some prospects for investigating these higher-order statistics.

\section{DFV Theory and High-Schmidt Limit}\label{S2}

In this section, we briefly review the work of DFV  \cite{donev2014reversible}
on diffusive mixing in the limit of large Schmidt number and low Mach number in an isothermal 
quiescent fluid. A more detailed survey can be found in \cite{EyinkJafari2022}. %The DFV theory is based on a formal adiabatic mode-elimination for the fast velocity degrees of freedom that yields reduced model equations for the scalar concentration field in the limit $Sc\gg 1$. Interestingly, the effective advecting velocity in these reduced equations for the scalar field turns out to be Gaussian and white-noise in time, i.e., the Kraichnan velocity.  
The starting point of the DFV theory differs from \eqref{Momentum100},\eqref{Passive110} in two 
important respects. The equation \eqref{Momentum100} for the fluctuating velocity field is 
unchanged, but DFV assumed that the concentration field $c({\bf x}, t)$ in a binary mixture 
satisfies the modified equation
\be\label{Passive111}
\partial_t c=-\bu'\bdot\grad c +\grad\bdot \left(D_0 \grad c\right).
\ee
This differs from \eqref{Passive110} firstly because $D_0$ now represents 
the {\it bare molecular diffusivity} before dressing by thermal fluctuations.  
Secondly, $\bu'$ is a {\it coarse-grained advection velocity} obtained by convolving
the solution $\bf v'$ of \eqref{Momentum100} with a smoothing kernel $\boldsymbol\sigma$, 
\begin{eqnarray}\label{regularized-velocity100}
\bu'(\bx, t)&\equiv& {\boldsymbol\sigma}\star{\bf v}'=\int {\boldsymbol\sigma}(\bx, \bx'){\bf v}'(\bx', t) \, d^3x'.
\end{eqnarray}
The cutoff length scale in this kernel, denoted $\sigma,$ is taken to be of the order of the 
radius of the solute molecules. The underlying assumption is that the molecules (or perhaps even colloidal 
particles in suspension) feel only the fluctuating velocity field averaged over their extent.  

With these basic assumptions, DFV then performed an exact asymptotic analysis in the limit of high Schmidt-number limit, $Sc_0\equiv \frac{\eta}{D_0 \rho}\gg 1.$ Motivated by the well-known Stokes-Einstein 
relation $D\sim k_B T/ \eta\sigma$, DFV introduced a small parameter $\epsilon\ll 1$ to order quantities 
for formal asymptotics, adopting the scaling 
\be \eta\mapsto \epsilon^{-1}\eta, \quad D_0\mapsto \epsilon D_0 \lb{trans-to} \ee
such that $D_0 \eta\simeq (const.)$ and $Sc_0\sim \epsilon^{-2}.$  In the limit $\epsilon\ll 1$, there exists a 
time scale separation between the fast viscous dynamics, governing the thermal velocity fluctuations ${\bf v}'$, 
and much slower diffusive evolution of the concentration field $c.$ To formalize this time separation, DFV introduced
a ``macroscopic'' diffusive time $\tau$ which is related to the ``microscopic''
viscous time $t$ of Eqs. \eqref{Momentum100},\eqref{Passive110} 
by $t=\epsilon^{-1} \tau,$ or, equivalently, by the scaling
\be t\mapsto \epsilon^{-1}t \lb{time-to} \ee 
with $\tau$ renamed $t$. This procedure leads to a limiting stochastic advection-diffusion
equation for the concentration field in the ``macroscopic'' time:
\be\label{PassiveStratonovich100}
\partial_t c=-{\bf w}\odot\grad c+D_0 \triangle c,
\ee
%\be\label{PassiveStratonovich100}
%\partial_t c=-{\bf w}\odot\grad c+D_0 \triangle c+\grad\bdot  \Big(\sqrt{2mD_0\rho^{-1} c(1-c)} \; {\boldsymbol\eta}_c \Big),
%\ee
where $\odot$ represents a Stratonovich dot product and ${\bf w}({\bf x}, t)$ is an incompressible, advecting 
random velocity field which is white noise in time, with zero mean and covariance 
\bea \label{SmoothVelocityCovariance101}
&& \langle {\bf w}({\bf x}, t) \otimes {\bf w}({\bf x'}, t') \rangle={ \cal {\bf R}} ({\bf x, x'}) \delta(t-t'), \cr 
&& { \cal {\bf R}} ({\bf x, x'})=\frac{2 k_B T}{\eta}(\bsigma\star\bG\star\bsigma^\top)(\bx, \bx').
\eea
The tensor $\bf G$ is the Green's function of the linear Stokes operator, which 
is singular for $\bx=\bx',$ unlike the smoothed tensor $\bR$ which is regular at coinciding points. Because of the delta-in-time correlation of the random velocity $\bw$, Eq.(\ref{PassiveStratonovich100}) for the concentration field is in fact a version of the Kraichnan model (see below).The spatial realizations of random thermal velocity $\bw$ can be obtained from the stationary Stokes equation with smoothed thermal forcing 
\bea \lb{wStokes}
-\grad q+ \nu\triangle\bw+\grad\bdot\left( \sqrt{\frac{2\nu k_BT}{\rho}} \boeta_\sigma \right)=\bzed
\eea 
where $\boeta_\sigma=\bsigma\star\boeta$ and $q$ is determined by $\grad\bdot\bw=0$. The physical content of the above equation is that viscous diffusion and thermal fluctuations are in instantaneous balance for the effective advecting velocity $\bw,$ with long-range spatial correlations induced by the incompressibility condition. 

Whereas the starting advection-diffusion equation in the DFV theory, Eq. (\ref{Passive111}), contains  the bare diffusivity $D_0$ the reduced equation (\ref{PassiveStratonovich100}) in the limit $Sc\gg 1$ involves a \textit{renormalized diffusivity}. This can be seen easily by writing the It$\baro$ form of Eq.(\ref{PassiveStratonovich100}):  
\begin{eqnarray}\label{PassiveIto100}
\partial_t c&=& -{\bf w}\bdot\grad c+\grad\bdot ({\bf D(x)}\grad c ),
\end{eqnarray}
%\begin{eqnarray}\nonumber
%\partial_t c&=& -{\bf w}\bdot\grad c+D_0 \triangle c+\grad\bdot ({\bf D(x)}\grad c )\\\label{PassiveIto100}
%&&   +\grad\bdot \Big(\sqrt{2mD_0\rho^{-1} c(1-c)} \; {\boldsymbol\eta}_c \Big),
%\end{eqnarray}
with
\be {\bf D(x)}:=D_0{\bf I}+\frac{1}{2} { \cal {\bf R}} ({\bf x, x}). \lb{Deff} \ee
The latter renormalized diffusivity originates from advection by the viscously slaved thermal velocity 
fluctuations and is similar to an ``eddy-diffusivity'' due to eliminated turbulent eddies in turbulent flows. 
As emphasized by DFV, generally $D_0\ll \abs{{\bf D}({\bf x})}$ and, indeed, one may consider the limit 
of vanishing bare diffusivity, $D_0\to 0,$ with essentially no empirical consequence.  

The original paper of DFV elaborated two types of numerical algorithm to solve for realizations of $c(\bx,t)$ in the asymptotic theory; see \cite{donev2014reversible}, Appendix B. The first Eulerian method 
was based on a straightforward predictor-corrector integrator for the stochastic advection-diffusion equation
\eqref{PassiveStratonovich100} with a staggered finite-volume discretization in space. The  realizations 
of ${\bf w}$ were obtained from the Stokes equation \eqref{wStokes} with an iterative Krylov linear solver. 
Discretizing the advection term ${\bf w}\odot\grad c$ with a non-dissipative centered-difference 
formula was shown to maintain discrete fluctuation-dissipation balance, but requires a non-vanishing bare diffusivity $D_0$ for numerical stability. A second Lagrangian numerical scheme was obtained by 
a spectral decomposition of the covariance 
$$ { \cal {\bf R}} ({\bf x, x}') = \sum_\alpha \bphi_\alpha(\bx)\bphi_\alpha(\bx') $$
and by then solving the Lagrangian equations for space positions of $N$ tracer particles 
\begin{eqnarray} 
&& d{\bf q}_n(t)/dt \,=\,  {\bf w}({\bf q}_n(t),t) + \sqrt{2D_0}\,\boeta_0(t), \quad n=1,...,N   \cr
&&\qquad \,=\, \sum_\alpha \bphi_\alpha({\bf q}_n(t)) \eta_\alpha(t) + \sqrt{2D_0}\,\boeta_0(t), 
\label{dQdt} 
\end{eqnarray} 
with i.i.d. scalar white-noises $\eta_\alpha(t)$ and vector white noise $\boeta_0(t).$ Efficient evaluation 
of the summation over $\alpha$ in \eqref{dQdt} requires a non-uniform FFT method to evaluate 
${\bf w}(\bx,t)$ only at the particle locations. Finally, 
with initial tracer particle positions distributed in space according to the concentration field
$c_0(\bx),$ then the field at later times is approximated by 
$$ c(\bx,t) = \frac{1}{N} \sum_{n=1}^N \delta^3(\bx-{\bf q}_n(t)). $$
This algorithm has cost scaling linearly in $N$ and can be applied with $D_0=0,$ but requires 
large numbers of particles $N$ for convergence. Both of these numerical schemes can be applied 
in the experimentally relevant case of finite spatial domains and can incorporate additional 
important effects, such as buoyancy. 

Another powerful feature of the DFV theory, although not fully exploited in \cite{donev2014reversible}, is the existence of closed partial differential equations for the scalar correlation functions of any arbitrary order. This follows from the fact that the long-time, high-$Sc$ limit concentration equation in the DFV theory, Eq.(\ref{PassiveStratonovich100}), is indeed a version of the Kraichnan model \cite{kraichnan1968small,kraichnan1974convection,
falkovich2001particles} in which there is no closure problem for correlation functions. %Indeed, the ``Kraichnan model'' can be described mathematically by an equation similar to Eq.(\ref{PassiveStratonovich100}) for a passive scalar field advected by a Gaussian random velocity field $\bw(\bx,t)$ which is white-noise in time, with zero mean and covariance given by an expression similar to Eq.(\ref{SmoothVelocityCovariance101}). The incompressiblity constraint, assumed in this paper, can be relaxed as the compressible Kraichnan model has also been studied in the literature
% \cite{gawedzki2002soluble,falkovich2001particles}.
In the Kraichnan model, the equal-time $p$-point correlation functions 
$C_p({\bf x}_1, \dots, {\bf x}_p;t):=\langle c({\bf x}_1, t)c({\bf x}_2, t)...c({\bf x}_p, t)\rangle$
averaged over realizations of the thermal velocity ${\bf w}$ for scalar field $c(\bx, t)$ governed by Eq.(\ref{PassiveStratonovich100}) satisfy an exact closed differential equation (see e.g., \cite{EyinkJafari2022}; for a detailed review of the Kraichnan model see \cite{falkovich2001particles, Gawedzki1997,gawedzki2002easy,gawedzki2002soluble,cardy2008non}). For $p=1,$
$C_1(\bx,t)=\langle c(\bx,t)\rangle:=\overline c(\bx, t),$ the mean concentration field, satisfies the diffusion equation with renormalized diffusivity:
\begin{eqnarray}\label{C1eq}
&& \partial_t \bar{c}({\bf x}, t)
={1\over 2} 
\nabla_{x^i}\left[{ R}_{ij}(\bx,\bx)\nabla_{x^j} \bar{c}({\bf x}, t) \right]
+D_0 \triangle_{{\bf x}} \bar{c}({\bf x}, t). 
\end{eqnarray}
and for $p=2$ 
\begin{eqnarray}\nonumber
&&{\partial\over\partial t} C_2({\bf x}_1,{\bf x}_2, t) =D_0 \sum_{n=1}^2 \triangle_{{\bf x}_n}  C_2({\bf x}_1,{\bf x}_2, t)\\\label{Kraichnan1}
&&\;+{1\over 2} \sum_{n, m=1}^2 
\nabla_{x_n^i}\left[{ R}_{ij}(\bx_n,\bx_m)\nabla_{x_m^j}  C_2({\bf x}_1,{\bf x}_2, t) \right],
\end{eqnarray}
where ${\bf R}$ is the spatial covariance of the random velocity field, given by Eq.(\ref{SmoothVelocityCovariance101}). Combining these two equations yields a corresponding 
closed PDE for the concentration 2nd-order cumulant (connected correlation) function ${\mathcal C}({\bf x}_1, {\bf x}_2, t):=  
\langle c({\bf x}_1,t) c({\bf x}_2, t)\rangle-\langle c({\bf x}_1,t)\rangle \langle c({\bf x}_2,t)\rangle=
C_2({\bf x}_1,{\bf x}_2, t)-\overline c({\bf x}_1,t)  \overline c({\bf x}_2,t)$: 
\begin{eqnarray}\nonumber
 &&\partial_t {\mathcal C}({\bf x}_1, {\bf x}_2, t)=D_0 \sum_{n=1}^2 \triangle_{{\bf x}_n} {\mathcal C}({\bf x}_1,{\bf x}_2, t)\\\nonumber
&&\;+{1\over 2} \sum_{n, m=1}^2 
\nabla_{x_n^i}\left[{ R}_{ij}(\bx_n,\bx_m)\nabla_{x_m^j} {\mathcal C}({\bf x}_1,{\bf x}_2, t) \right]\\\label{C2eq}
&&\; + R_{ij}({\bf x}_1,{\bf x}_2) \nabla_{x_1^i}\bar{c}({\bf x}_1,t)\nabla_{x_2^j}\bar{c}({\bf x}_2,t)
\end{eqnarray}
Similar results hold for all $p>2.$

%Different techniques have been employed in the context of the Kraichnan model to study higher-order correlations \cite{falkovich2001particles,Gawedzki1997,gawedzki2002easy,gawedzki2002soluble} and even individual realizations of the concentration field \cite{gawedzki2002soluble,lototskii2004passive}. 

Furthermore, multi-time concentration correlations such as ${\mathcal C}(\bx,t;\bx',t'):=
\langle c(\bx, t)c(\bx',t')\rangle-\langle c(\bx, t)\rangle\langle c(\bx',t')\rangle$
can also be obtained in the Kraichnan model. Once the single-time time correlations are known, 
these can be obtained by solving additional closed PDE's, e.g. for $p=2$ one must solve 
the equation 
\bea\label{Cmultitime} 
&& \partial_t {\mathcal C}(\bx,t;\bx',t')\cr 
&& \hspace{10pt} 
= \grad_\bx\bdot\left(\left(D_0{\bf I}+\frac{1}{2}\bR(\bx,\bx)\right)\bdot\grad_\bx 
{\mathcal C}(\bx,t;\bx',t')\right) 
\hspace{10pt} 
\eea
for $t>t'$ with ${\mathcal C}(\bx,\bx',t')$ as initial data. The equation (\ref{Cmultitime}) 
follows because of the Markovian character of the dynamics in the It$\bar{{\rm o}}$ form
\eqref{PassiveIto100}. Clearly, this equation implies that the correlations of concentration 
fluctuations in the Kraichnan model are relaxed by diffusion with the renormalized diffusivity.
Thus, the Onsager regression hypothesis \cite{onsager1931reciprocalI,onsager1931reciprocalII} 
becomes exact in the high-Schmidt limit instantaneously in time, without the need for any aging interval.  Because the Kraichnan model dynamics is 
time-reversible, the results for $t<t'$ are identical to those for $t>t',$ 
depending only upon $\lvert t-t'\rvert$. 

The above closed equations have the form of parabolic diffusion equations with position-dependent 
diffusivities. They can be derived in finite domains with realistic boundary conditions, 
where they may provide valuable predictions for experimental flows. In general,
these equations must be solved computationally using numerical PDE methods, which will 
become more challenging for complex geometries and especially for high-order statistics 
with $p>2$ when the diffusion equation is very multi-dimensional. In the next section we discuss
a simple setting where exact analytical solutions are possible.

\section{Diffusion in an Unbounded Fluid}\label{sec:unbd} 

To get some insight about the predictions of the DFV theory, it is useful to consider 
the idealized case of a fluid mixture in unbounded 3D space, so that velocity statistics 
are spatially homogeneous and isotropic. In that case the Green's function of the 
Stokes operator is given also in closed form by the Oseen-Burgers tensor  
\be G_{ij}(\bx, \bx')=G_{ij}(\bx- \bx')={1\over 8\pi r}(\delta_{ij}+{r_ir_j\over r^2}) \lb{oseen} \ee 
with $\bf r=\bf x-\bf x'.$ Consequently, the covariance of the random thermal velocity ${\bf w}$
in the DFV theory becomes 
\begin{equation}\label{R}
R_{ij}({\bf r})={k_BT\over 4\pi \eta r}\Big(\delta_{ij}+{r_i r_j\over r^2}\Big), \quad r\gg \sigma
\end{equation}
independent of the choice of the filter kernel $\bsigma$ in \eqref{SmoothVelocityCovariance101}. 
In a homogeneous and isotropic flow of this type, the effective diffusivity $\bD$ defined in \eqref{Deff} becomes independent of $\bx$  with $D_{ij}=R_{ij}(0)/2=D\delta_{ij}.$ (Here and hereafter
we assume $D_0=0$ for simplicity). The precise value of $D$ does depend now upon the filter kernel
$\bsigma,$ with the kernel used in \cite{donev2014reversible} selected precisely to give 
\be D={k_BT\over 6\pi \eta\sigma} \lb{SE} \ee 
in agreement with the Stokes-Einstein relation for a suspension of hard spheres of radius $\sigma.$
However, any reasonable choice of filter gives a result identical to \eqref{SE} with $6\pi$
replaced by another numerical constant of order unity. Thus, with a suitable redefinition of the filter width $\sigma,$ sometimes called the ``hydrodynamic radius,'' the result in \eqref{SE} can always be taken to hold by convention. 
The DFV theory thus explains the empirical success of the Stokes-Einstein formula as the effect of strong renormalization of a small bare diffusivity due to advection of solute molecules by thermal 
velocity fluctuations. 

We next discuss some specific situations in unbounded domains where the DFV closed equations 
for mean and correlation function of the concentration field can be simplified and even solved 
analytically. 

\subsection{Free Diffusion}

%As discussed in Sec.\ref{S2}, striking experimental evidence supports the existence of giant concentration fluctuations,  associated with structure function $S(k)\sim k^{-4}$, in transient decay \cite{vailati1997giant,croccolo2007nondiffusive}. Theoretically, this problem has been approached using linearized fluctuating hydrodynamics \cite{vailati1998nonequilibrium}, however, systematic deviations are observed between linearized theory and experiment at early times when concentration gradients are very large: see \cite{croccolo2007nondiffusive},
%Figure 8. Thus it is interesting to study free decay using the DFV's fully non-linear theory to see whether its predictions differ from, or confirm, those of the linearized theory.

As discussed in the Introduction, giant concentration fluctuations with 
structure function $S(k,t)\propto k^{-4}$ have been observed very notably in free decay 
experiments where an initial blob of concentration diffuses in a fluid at rest
\cite{vailati1997giant,brogioli2000universal,croccolo2007nondiffusive}.
We illustrate here the closed equations that arise from the DFV theory in such free diffusion 
when the fluid is idealised as unbounded. 

\begin{figure}
\includegraphics[scale=.25]{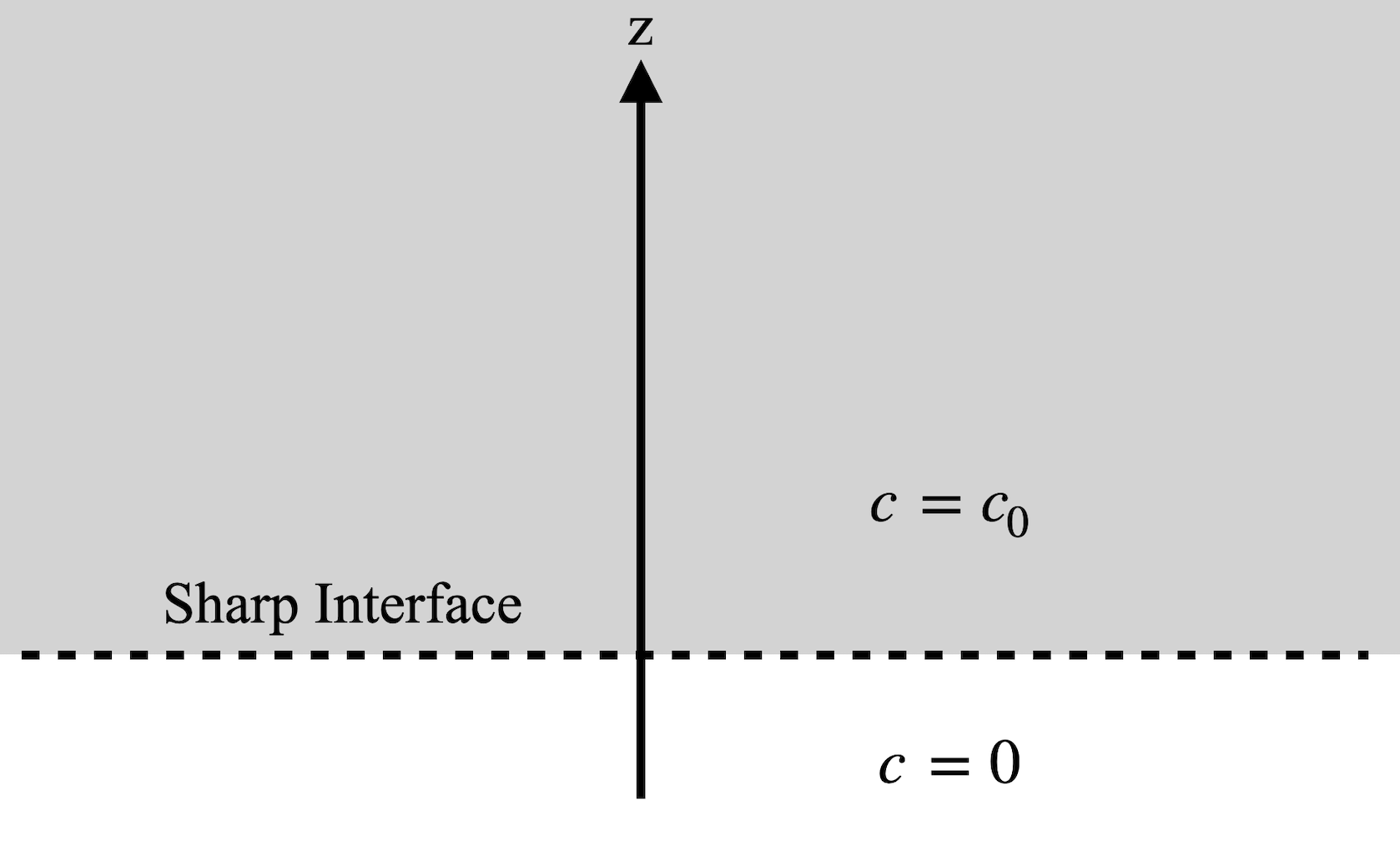}
\centering
\caption {\footnotesize {Initial distribution $(t=0)$ for a concentration field $c(z, t)$ with non-zero value $c=c_0$ in the region $z>0$ with a sharp interface at $z=0$.}} %Note the isotropy on the $xy$ plane perpendicular to the $z$-axis. }}
\label{configuration}
\end{figure}

A simple initial configuration of the scalar with maximal symmetry is a plane sharp interface at $z=0$, dividing two regions $z>0$ with constant concentrations 
\begin{equation} c = \left\{\begin{array}{ll}
               c_0 & z>0 \cr
               0 & z<0   \cr
               \end{array} \right. \end{equation} 
See Fig.\ref{configuration}. We consider here deterministic 
initial data for the concentration field, for simplicity, but it is easy to 
include thermal equilibrium fluctuations or other zero-mean random perturbations
(see below). The equation \eqref{C1eq} for the mean concentration profile 
reduces here to the 1D diffusion equation,
\begin{equation}\label{Diff1}
{\partial\over \partial t} \overline c(z, t)=D\partial^2_z\overline c(z, t)
\end{equation} 
with diffusivity $D$ given by \eqref{SE}. For the step-function initial condition, 
this has the well-known solution 

\bea
\overline c (z, t)={c_0\over 2}\Big(1+\erf\Big({z\over 2\sqrt{Dt}}\Big) \Big),
\lb{erfc} \eea 
in terms of the error function $\erf(z).$  

The equation \eqref{C2eq} for the 2nd-order cumulant can be simplified by introducing 
relative and mean variables
$$ {\bf r}={\bf x}_1-{\bf x}_2=(x,y,z), \quad {\bf X}=\frac{1}{2}({\bf x}_1+{\bf x}_2)=(X,Y,Z) $$
and by noting that, from 2D translational symmetry, ${\mathcal C}$ depends only upon ${\bf r}$ and $Z,$
so that: 
\begin{eqnarray}\nonumber
\partial_t {\mathcal C}&=&{1\over 2}\Big(D+{R_{33}({\bf r})\over 2}\Big)\partial_Z^2 {\mathcal C}+\Big(R_{ij}({\bf 0})-R_{ij}({\bf r})\Big)\partial_i\partial_j {\mathcal C}\\\label{C_2}
&&+R_{33}({\bf r})\nabla\overline c\Big(Z+{z\over 2}\Big)\nabla\overline c \Big(Z-{z\over 2}\Big). \label{C2unbd} 
\end{eqnarray}
However, furthermore, there is rotational symmetry around the $z$-axis. 
%Working in the cylindrical coordinates $(\rho, \phi, z)\equiv (r_\perp, \phi, r_\parallel)$, this 
%means that ${\mathcal C}={\mathcal C}(r_\|,r_\perp,Z,t)$ must be independent of $\phi.$
%Using Eq.(\ref{R}) for the covariance tensor $R_{ij}({\bf r})$, it is then straightforward 
%to show that \eqref{C2unbd} becomes for $r\gg \sigma$
%\begin{eqnarray}\nonumber
%\partial_t {\mathcal C} 
%&=& {k_BT\over 4\pi\eta\sigma}\left(\frac{1}{3}+\frac{\sigma}{4r}\left(1+\frac{r_\|^2}{r^2}\right)\right)
%\partial_Z^2 {\mathcal C}\\\nonumber
%&&+{k_BT\over 3\pi\eta \sigma}\Big( 1-{3\over 4}{\sigma\over r} \Big)\Big({1\over r_\perp}\partial_\perp +\partial_\perp^2+\partial_\parallel^2 \Big){\mathcal C}\\\nonumber
%&&-{k_BT\over 4\pi\eta}{1\over r^3 }\Big( r_\parallel^2\partial_\parallel^2 +r_\perp^2\partial_\perp^2 +2r_\perp r_\parallel \partial^2_{\perp,\parallel}\Big){\mathcal C}\\\label{Cspherical}
%&&+{k_BT\over 4\pi\eta }{1\over r} \Big(1+{r_\parallel^2 \over r^2} \Big)
%\grad\overline c\Big(Z+{z\over 2}\Big)\grad \overline c \Big(Z-{z\over 2}\Big),\qquad
%\end{eqnarray}
%where we have adopted the notations $r=(r^2_\|+r^2_\perp)^{1/2},$
%$\partial_\parallel:={\partial\over\partial r_\parallel}$, $\partial_\perp:={\partial\over\partial r_\perp}$ and $\partial^2_{\perp,\parallel}:={\partial^2\over\partial r_\perp\partial_\parallel}$. 
%Eq.(\ref{Cspherical}) can be transformed to an even simpler form by using the spherical coordinates 
%defined as $(r, \theta, \phi)$, see Fig.(\ref{zap1}), so that  $r_\perp=r \sin\theta$ and $r_\parallel=r \cos\theta.$ 
Working in spherical coordinates defined as $(r, \theta, \phi)$, 
%see Fig.~\ref{zap1}, 
this means that ${\mathcal C}={\mathcal C}(r,\theta,Z,t)$ must be independent of azimuthal angle $\phi.$ In those variables, it is tedious but 
straightforward to show that \eqref{C2unbd} becomes for $r\gg \sigma$
\begin{eqnarray}\nonumber
\partial_t {\mathcal C} 
&=& {k_BT\over 4\pi\eta\sigma}\left(\frac{1}{3}+\frac{\sigma}{4r}(1+\cos^2\theta)\right)
\partial_Z^2 {\mathcal C}\\\nonumber
&&+{k_BT\over 3\pi\eta\sigma}\cdot {1\over r^2}{\partial\over\partial r}\Big[ r^2\Big(1-{3\over 2}{\sigma\over r}\Big){\partial {\mathcal C}\over\partial r} \Big]
\\\nonumber
&&+{k_BT\over 3\pi\eta\sigma}\cdot {1\over r^2\sin\theta}\Big(1-{3\over 4}{\sigma\over r}\Big){\partial\over\partial\theta}\Big(\sin\theta{\partial {\mathcal C}\over\partial\theta}  \Big)\\\label{anisotropic1}
&&+ {k_BT\over 4\pi\eta} \cdot {1\over r}(1+\cos^2\theta)
\nabla\overline c\Big(Z+{z\over 2}\Big)\nabla \overline c \Big(Z-{z\over 2}\Big),\qquad
\label{Ceq-slab} \end{eqnarray}
where $r=(x^2+y^2+z^2)^{1/2}$ and $z=r \cos\theta.$ 
%Because of anisotropy, imposed by the concentration gradient which corresponds to the $\theta$-dependent terms, finding a solution to this equation is not trivial.
Here we can recognize the scale-dependent diffusivity 
 \bea\label{D}
 D(r)={k_BT\over 6\pi \eta \sigma}\Big(1-{3\over 2}{\sigma\over r}\Big).
 \eea
that appeared in our earlier work \cite{EyinkJafari2022}, Eq.(58), for the fully isotropic problem.  

The end result \eqref{Ceq-slab} is a diffusion equation in the 3D space of variables 
$(r,\theta,Z).$ The initial correlation ${\mathcal C}_0$ will depend upon the 
random initial concentration field adopted, e.g. ${\mathcal C}_0\equiv 0$ 
for the deterministic initial data discussed above. Even for the highly symmetric 
physical configuration considered, the simplified version of the general equation 
\eqref{C2eq} for the concentration correlation ${\mathcal C}$
is too complicated for a fully analytical solution. However, with proper regularization 
of the divergences at $r=0$ (e.g. see \cite{EyinkJafari2022}, Eq.(56)-(57)),  the PDE \eqref{Ceq-slab} 
can be solved by numerical discretization methods. Furthermore, the solution may be rationally 
approximated by a combination  of numerics and analysis. Another case which yields 
similar simplifications is the problem of an initial spherical blob of concentration. 
Nonequilibrium concentration fluctuations in inhomogeneous, anisotropic and time-dependent free decay  
is the subject of ongoing work based on the DFV theory and will be presented elsewhere. 
However, it turns out that with one further simplification the equation \eqref{Ceq-slab}
becomes mathematically solvable, as discussed in the following subsection.

\subsection{Constant Concentration Gradient} 

An even simpler limiting case of the problem in the previous section is obtained 
by letting $t\to\infty,$ $c_0\to\infty$ so that $t=c_0^2/4D\pi \gamma^2$ is held 
fixed for some constant $\gamma.$ In that case the error-function profile \eqref{erfc},
after subtraction of the divergent constant $c_0,$ reduces to the case of constant 
gradient, $\bar{c}=(\nabla \bar{c})z$ with 
$\nabla\bar{c}=\gamma.$ This idealized situation has been studied before 
using linearized fluctuating hydrodynamics and exactly solved within that approximation 
\cite{vailati1998nonequilibrium,brogioli2000diffusive,brogioli2016correlations}.
 Here we study this same problem using the high-Schmidt asymptotic theory of DFV
 without neglecting nonlinear interactions. 

This problem remains anisotropic because of the imposed gradient but it is now fully 
space-homogeneous and yields a time-independent steady-state. Therefore we may neglect
derivatives with respect to $Z$ and $t$ in equation \eqref{Ceq-slab}, yielding  
\begin{eqnarray}\nonumber
&&{2\over r^2}{\partial\over\partial r}\Big[ r^2\Big(1-{3\over 2}{\sigma\over r}\Big){\partial {\mathcal C}\over\partial r} \Big] +{2\over r^2\sin\theta}\Big(1-{3\over 4}{\sigma\over r}\Big){\partial\over\partial\theta}\Big(\sin\theta{\partial {\mathcal C}\over\partial\theta}  \Big)\\\label{anisotropic2}
&&\quad +{3\over 2}{\sigma\over r}(1+\cos^2\theta)\abs{\nabla \overline c }^2=0,
\lb{Cgrad} \end{eqnarray}
whose solution is the steady-state correlation ${\mathcal C}(r,\theta),$

As a first step in finding the solution, we may consider the correlation function 
averaged over solid angle: 
\begin{equation}
\overline{{\mathcal C}} (r):={1\over 4\pi} \int {\mathcal C}(r, \theta) d\Omega={1\over 2}\int_0^\pi \sin\theta \; {\mathcal C}(r, \theta) d\theta.
\lb{Cavrg} \end{equation}
Because of linearity of the equation \eqref{Cgrad}, this averaged correlation satisfies the ODE
\begin{eqnarray}\label{anisotropic2}
{1\over r^2}{\partial\over\partial r}\Big[ r^2\Big(1-{3\over 2}{\sigma\over r}\Big){\partial \overline {{\mathcal C}}\over\partial r} \Big]+{\sigma\over r}\abs{\nabla \overline c }^2=0.
\end{eqnarray}
Physically, this ODE corresponds to an isotropic system where the concentration gradient 
is not pointing in the $z$-coordinate direction but is instead pointing in a random 
direction uniformly distributed over solid angles. The general solution of this ODE is of the form 
\begin{eqnarray}\label{isotropic1}
\overline{{\mathcal C}}(r)
&=&-{\sigma \abs{\nabla \overline c }^2\over 2}\Big(r+{3\over 2}\sigma \ln\abs{ r-\frac{3}{2}\sigma}\Big) + A \ln\abs{1-\frac{3\sigma}{2r}}+B
\end{eqnarray}
for constants $A,B,$ but in the large-$r$ range of interest the solution reduces to 
$$ \overline{{\mathcal C}}(r) \doteq -{\sigma \abs{\nabla \overline c }^2\over 2}r + {\rm const.}, 
\quad r\gg \sigma. $$
This result shows the expected scaling $\propto r$ in physical space associated to GCF's
\cite{brogioli2016correlations,eyink2022high}. 

We can now look for the solution of the original anisotropic problem \eqref{Cgrad} in the separable 
form ${\mathcal C}(r, \theta)= \overline{{\mathcal C}}(r) {\mathcal G}(\theta).$ This strategy is successful for the range $r\gg\sigma$
and leads to the following autonomous equation for ${\mathcal G}$
\begin{equation}\label{G1}
{1\over\sin\theta}
{\partial \over \partial \theta}\left(\sin\theta {\partial {\mathcal G}\over\partial \theta}\right)
+2{\mathcal G}= {3\over 2}(1+\cos^2\theta).
\end{equation}
This ODE can be solved easily by first substituting $\xi=\cos\theta$ and finding a particular solution and general homogeneous solution by the Frobenius series method. One of the homogeneous solutions 
contains a logarithm and must be discarded as unphysical due to its non-smoothness at $\theta=\pi/2$.
The result for the anisotropy factor is then found to be  
\begin{eqnarray}\label{G}
{\mathcal G}(\theta)&=& {3\over 8} (2+\sin^2\theta)+A\Big(1-{\cos^2\theta\over 2}-{\cos^4\theta\over 8}+\dots \Big),
\end{eqnarray}
where the constant $A$ is to be determined by the condition \eqref{Cavrg}. It turns out 
that the latter condition is satisfied exactly by the indicated particular solution, so that 
$A=0$ and we obtain the final result
\begin{equation}\label{anisotropic20}
{\mathcal C}(r, \theta)\simeq -{3\over 16} \sigma \abs{\nabla \overline c }^2\;r (2+\sin^2\theta),
\quad r\gg \sigma.
\end{equation}
To our knowledge, this full physical-space correlation of GCF's has not been noted in the literature before. See \cite{brogioli2016correlations} for previous partial results. 

The scaling of GCF's is more typically given in Fourier space by the {\it static structure function} 
\begin{eqnarray}\label{Structure1}
S(k)&=& \int {\mathcal C}(r, \theta) e^{i{\bf k}\bdot{\bf r}} d^3r\cr 
&=& \int_0^\infty r^2 dr \int_0^\pi \sin\theta d\theta \int_0^{2\pi} d\phi\; {\mathcal C}(r, \theta)e^{i(k_\parallel r \cos\theta +k_\perp r \sin\theta \cos\phi)},
\end{eqnarray}
where $k_\|$ is magnitude of the wavenumber component parallel to $\grad\bar{c}$ and 
$k_\perp$ is magnitude of the perpendicular component.
The argument of the exponential function in the above expression can be simplified by applying a rotation to the spherical coordinate system $(r, \theta,\phi)$ by angle $\alpha=\tan^{-1}(k_\perp/k_\parallel)$, which results in standard integrals over the new polar angle $\theta'$. Introducing an IR cut-off,  the integral over $r$ can be easily evaluated, with the final result given by 
\begin{eqnarray}\label{structure2}
S(k)=6\pi \sigma \abs{\nabla \overline c }^2 {{\hat k}_\perp^2\over k^4}
=\frac{k_B T}{D\eta} \abs{\nabla \overline c }^2 {{\hat k}_\perp^2\over k^4},  
\quad k\sigma\ll 1\quad 
\end{eqnarray}
where $\hat k_\perp$ is the perpendicular component of the unit vector $\hat{\bk}=\bk/k$. 
See Appendix \ref{Fourier}. Remarkably, 
this is identical to the result predicted by linearized fluctating hydrodynamics
\cite{vailati1998nonequilibrium,brogioli2000diffusive,brogioli2016correlations},
in the limit of high Schmidt number and with negligible bouyancy effects. 

The {\it dynamic structure function} is similarly defined in terms of the 
2-time correlation function ${\mathcal C}({\bf r},\tau)=
 \langle c'({\bf x}+{\bf r}, t+\tau) c'({\bf x}, t)\rangle$
 as 
\be S(k,\omega) =  \iint {\mathcal C}({\bf r},\tau) e^{i{\bf k}\bdot{\bf r}-i\omega\tau} \,d^3r \,d\tau. \ee 
Because of space-homogeneity the general equation \eqref{Cmultitime} for 2-time correlations 
reduces to 
\begin{equation}
\partial_\tau {\mathcal C}({\bf r}, \tau)=D \triangle_{\bf r} {\mathcal C}({\bf r},\tau)
\end{equation}
for $\tau>0$ and the corresponding anti-diffusion equation for $\tau<0.$
We thus obtain 
\begin{eqnarray}\nonumber
S(k, \omega) &=& \int_{-\infty}^{+\infty} S(k) e^{-k^2 D \abs{\tau}} e^{-i\omega \tau} d\tau\cr
&=&S(k){2D k^2 \over \omega^2+(Dk^2)^2} \cr 
&=& \frac{k_B T}{D\eta} \abs{\nabla \overline c }^2 {{\hat k}_\perp^2\over k^4}{2D k^2 \over \omega^2+(Dk^2)^2},
\quad k\sigma\ll 1 \label{Dynamic}
\end{eqnarray}
once again in perfect agreement with the prediction of linearized theory 
\cite{vailati1998nonequilibrium,brogioli2000diffusive,brogioli2016correlations}. 
Evident here is the central Rayleigh peak which determines the light scattering properties of the mixture.

\section{Conclusions}

It is not obvious why the DFV theory results \eqref{structure2},\eqref{Dynamic} 
for the concentration structure function should agree with the predictions of linearized fluctuating 
hydrodynamics in the limit of high Schmidt number. This agreement means that the neglected nonlinear 
terms which dynamically couple the fluctuations of velocity and concentration do not modify 
at all the 2nd-order correlations created by the direct random advection of the mean concentration 
gradient. The situation is somewhat reminiscent of the non-renormalization theorems for the chiral 
anomaly in quantum gauge theory. The Adler-Bardeeen theorem \cite{adler1969absence,adler2005anomalies}, 
nicely rederived by a renormalization group argument of Zee \cite{zee1972axial}, implies that the result 
of the leading-order triangle diagram is not modified by higher corrections from nonlinear interactions to 
any order in perturbation theory. This analogy is fitting for the scalar 2-point correlation function, 
since its scaling in the Kraichan model is well-known to be connected with the scalar dissipation 
anomaly. As discussed by Gaw\c{e}dzki  \cite{Gawedzki1997}, Lecture 3, or \cite{gawedzki2002easy},
Lecture 4, the 2-point correlation scaling in the Kraichnan model is analogous to the 
exact $4/5$th law in hydrodynamic turbulence which expresses the dissipative anomaly. Just as for 
the 3rd-order velocity structure function in hydrodynamic turbulence, the 2nd-order correlation 
in the Kraichnan model has scaling given by dimensional analysis and is protected from 
acquiring anomalous scaling because of the lack of non-constant zero modes of its 
linear evolution operator. But note that a true dissipative anomaly as $D_0\to 0$ will not occur 
for the concentration variance in diffusive mixing according to the DFV theory, due to the regularity of the 
advecting velocity ${\bf w}$ at scale $\sigma$! See \cite{donev2014reversible}, section 3.1.

The DFV theory results help to explain the remarkable agreement of linearized fluctuating hydrodynamics 
predictions with existing observations from experiment and simulation. They furthermore 
confirm the finding of \cite{eyink2022high} that the DFV theory reproduces the expected 
scaling $\propto k^{-4}$ of the static structure function. This result opens the door to the 
use of the DFV theory for problems of current interest with large gradients, high concentrations 
and transient dynamics \cite{baaske2016neuf,vailati2020giant}. Note that in the derivation 
of the equation \eqref{C1eq} for the mean concentration $\bar{c}({\bf x},t)$ and equation 
\eqref{C2eq} for the fluctuation correlation ${\mathcal C}({\bf x},{\bf x}',t)$ there 
is no assumption of a separation in space and time scales for these two quantities, which 
co-evolve together. Thus, joint solution of these two equations provides a more systematic approach
than current approximate solutions of the linearized equations that assume scale separation 
\cite{vailati1998nonequilibrium}. There is also no assumption of small concentrations or 
weak gradients in the derivation of the equations \eqref{C1eq},\eqref{C2eq}, 
which can be obtained as well with realistic boundary conditions in finite domains. For 
highly symmetric geometries such as the plane interface in an unbounded fluid discussed 
in section \ref{sec:unbd} these equations take simplified forms such as 
\eqref{Diff1},\eqref{Ceq-slab} which can be readily solved numerically, if not analytically. 
A Soret flux $\bJ_T=-D c S_T \grad \bar{T}$ associated with a mean temperature gradient 
can also be incorporated, although closure requires the assumption of weak concentrations so that 
$c(1-c)\simeq c.$ Without making any of these simplifying assumptions, the 
Eulerian and Lagrangian numerical methods originally developed in \cite{donev2014reversible} can be applied 
in finite domains and include additional important effects such as bouyancy. It is hoped that 
our analytical results will help to spur renewed interest in DFV's numerical methods,
since we now can have confidence that they yield physically valid results.

Lastly, but not least importantly, the DFV theory provides results also for higher 
$p$-point correlations of the concentration field with $p>2$ and our analysis gives no reason 
to believe that these will agree with the predictions of linearized fluctuating hydrodynamics. 
In particular, these higher-order statistics must certainly be non-Gaussian due to the 
nonlinear coupling term $-\bv'\bdot \grad c'$ neglected in linearized theory. Methods 
previously developed to study anomalous scaling in the Kraichnan model apply here, 
including analytical approaches such as $1/d$ expansion \cite{chertkov1995normal,chertkov1996anomalous}
and Lagrangian numerics \cite{frisch1998intermittency,gat1998anomalous,frisch1999lagrangian}.
We are currently investigating such higher-order correlations. These non-Gaussian statistics
are not only of theoretical interest but might also be observable experimentally, e.g. by 
multiple-scattering measurements \cite{lemieux1999investigating}. Such higher-order correlations
are now accessible in other condensed matter systems such as atomic quantum gases 
\cite{schweigler2017experimental} and, as in those cases, experimental access to such 
non-Gaussian statistics could provide an entirely new window into the non-equilibrium 
physics of diffusive mixing.

\backmatter

\bmhead{Acknowledgments}
We dedicate this paper to the memory of Krzysztof Gaw\c{e}dzki, to whom we are deeply indebted for his profound insights, his unfailing support and, not least, his warm 
friendship. We are also grateful to A. Vailati and A. Donev for stimulating discussions of nonequilibrium concentration fluctuations. 
We thank finally the Simons Foundation for support of this work with Targeted Grant No. MPS-663054, “Revisiting the Turbulence Problem Using Statistical Mechanics.”

\begin{appendices}

\section{Calculation of Structure Function}\label{Fourier}

\begin{figure}
\includegraphics[scale=.25]{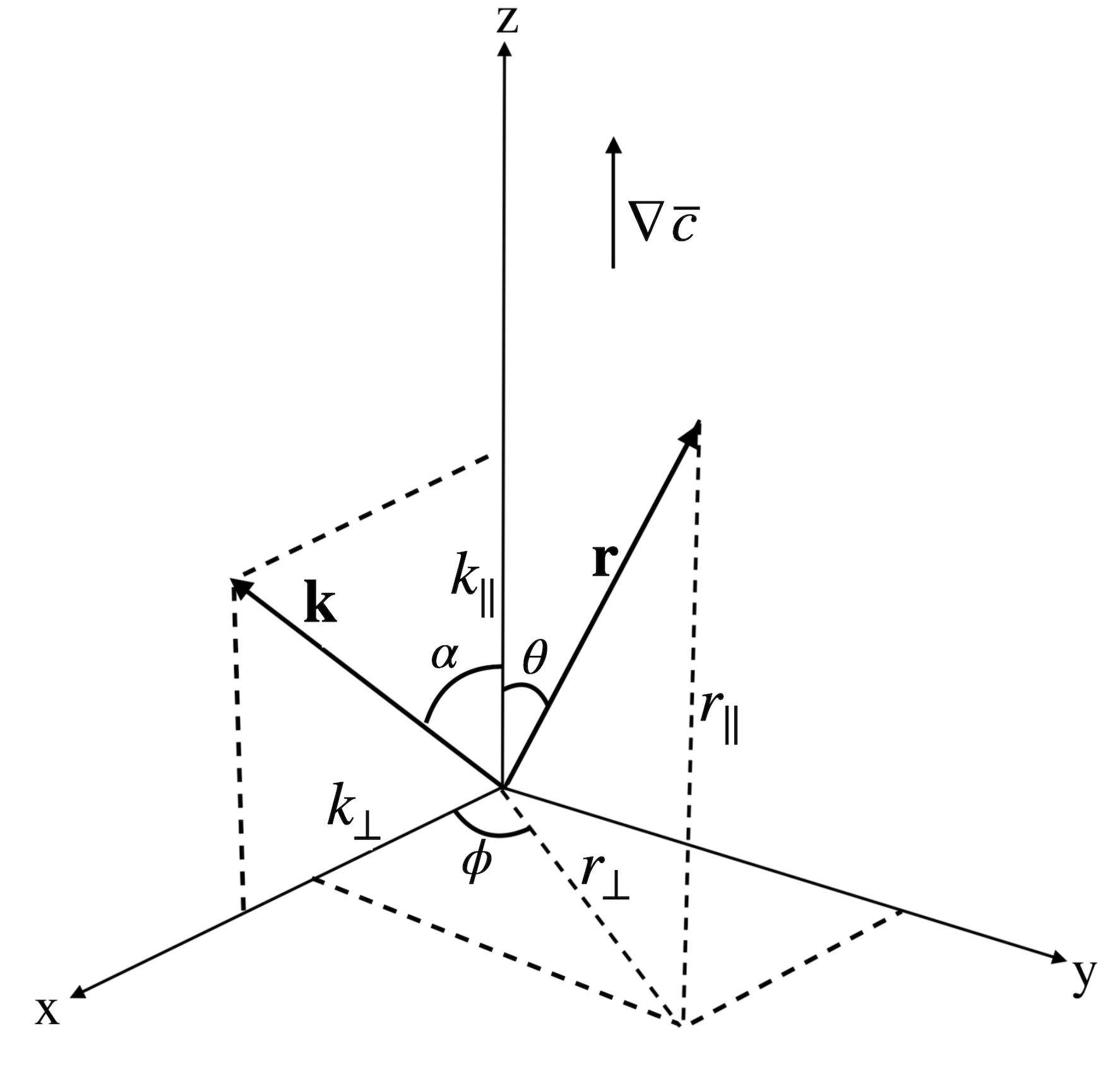}
\centering
\caption {\footnotesize {The spherical coordinate system used in evaluating  Eq.(\ref{Structure1}).}} %The expression $\cos(k_\parallel r \cos\theta +k_\perp r \sin\theta \cos\phi)\equiv \cos(k_\parallel r_\parallel+{\bf k}_\perp\bdot{\bf r}_\perp)$ in Eq.(\ref{Structure1}) can be greatly simplified by rotating the coordinate system by angle $\alpha=\tan^{-1}({k_\perp/ k_\parallel})$ such that in the new coordinates $\bf k$ becomes aligned with the $z$-axis.}}
\label{Rotation}
\end{figure}

In this appendix, we evaluate the Fourier transform of the correlation function ${\mathcal C}({\bf r})$ in Eq.(\ref{Structure1}) in order to obtain the static structure function $S(\bk)$. 
%The real part of Eq.(\ref{Structure1}) is\begin{eqnarray}\nonumberS(k)&=& \int_0^\infty r^2 dr \int_0^\pi \sin\theta d\theta \int_0^{2\pi} d\phi \\\nonumber&&\times \; C(r, \theta)\cos\Big(k_\parallel r \cos\theta +k_\perp r \sin\theta \cos\phi\Big)\\\nonumber&=& -{3\over 16} \sigma |\grad \langle c \rangle |^2 \int_0^\infty r^3 dr \int_0^\pi \sin\theta d\theta \int_0^{2\pi} d\phi \\\nonumber&&\times \; (2+\sin^2\theta)\cos\Big(k_\parallel r \cos\theta +k_\perp r \sin\theta \cos\phi\Big),\end{eqnarray}
First, we take the real part so that our integrand contains the expression $\cos({\bf k\cdot r})=\cos(k_\parallel r \cos\theta +k_\perp r \sin\theta \cos\phi)$ instead of the complex exponential. 
We next simplify the integral by rotating the $z$-axis of the integration variable
${\bf r}$ to be parallel with wavevector ${\bf k}$; see Fig.~\ref{Rotation}. This rotation 
by angle $\alpha=\tan^{-1}(k_\perp/k_\parallel)$ results in new coordinates, $(r, \theta', \phi')$, in which the argument $k_\parallel r \cos\theta +k_\perp r \sin\theta \cos\phi$ is transformed to $rk\cos\theta'$. The measure $d\Omega=\sin\theta d\theta d\phi=\sin\theta' d\theta' d\phi'$ is of course invariant under rotations. The remaining term in the integrand, $\sin^2\theta$, transforms as
$$\sin^2\theta\equiv (\cos\alpha\cos\phi'\sin\theta'+\sin\alpha\cos\theta')^2+\sin^2\phi'\sin^2\theta'$$
These simple results can easily be verified by applying the rotation matrix to an arbitrary unit vector $\hat u:=(\cos\phi\sin\theta, \sin\phi\sin\theta, \cos\theta)$:\begin{equation}\nonumber\begin{pmatrix}\cos\phi\sin\theta \\\sin\phi\sin\theta \\\cos\theta \end{pmatrix}=  \begin{pmatrix}\cos\alpha & 0 & \sin\alpha\\0 & 1 & 0\\-\sin\alpha & 0 & \cos\alpha\end{pmatrix}  \begin{pmatrix}\cos\phi'\sin\theta' \\\sin\phi'\sin\theta' \\\cos\theta' \end{pmatrix}.\end{equation}
We find
\begin{eqnarray}\nonumber
S(k)&=& -{3\over 16} \sigma \abs{\nabla \overline c}^2 \int_0^\infty r^3 dr \int_0^\pi \sin\theta d\theta \int_0^{2\pi} d\phi \\\nonumber
&&\times \; \Big(2+(\cos\alpha\cos\phi\sin\theta+\sin\alpha\cos\theta)^2+\sin^2\phi\sin^2\theta\Big)\cos(k r \cos\theta)
\end{eqnarray}
$\!\!\!$where, to simplify the notations, we have renamed the new primed coordinates $\theta', \phi'$ again as   $\theta, \phi$. We find
\begin{eqnarray}\nonumber
S(k)&=& -{3\over 8} \sigma \abs{\nabla \overline c}^2 \int_0^\infty e^{-r/L}r^3 dr \int_0^\pi \sin\theta d\theta \cos(k r \cos\theta)\int_0^{2\pi} d\phi\\
&&-{3\over 16} \sigma \abs{\nabla \overline c}^2 \int_0^\infty e^{-r/L} r^3 dr \int_0^\pi \sin\theta \cos(kr\cos\theta) d\theta \int_0^{2\pi} d\phi \\\nonumber
&&\times\Big(\cos^2\alpha\cos^2\phi\sin^2\theta+\sin^2\alpha\cos^2\theta+\sin 2\alpha\sin \theta\cos\theta+\sin^2\phi\sin^2\theta\Big)\\\nonumber\label{C10}
\end{eqnarray}

\vspace{-20pt} 
\noindent
where we have also introduced an IR cut-off $L$, with the understanding that the limit $L\to \infty$ will be taken at the end. 
The double integral in the first line of the above expression is straightforward to evaluate: first note that $\int_0^\pi d\theta \sin\theta \cos(k r \cos\theta)={2\over kr }\sin(kr)$, which can be integrated over $r$ using standard tables of integrals. For example, using \cite{Erdelyi1954} formula 4.7 (14), we find $\lim_{L\to +\infty}\int_0^\infty e^{-r/L} (kr)^2 \sin(kr) dr={-2\over k}$. Therefore, the first line of Eq.(\ref{C10}) reads
\begin{eqnarray}\label{1st}
&&-{3\pi \over 4} \sigma \abs{\nabla  \overline c }^2 \int_0^\infty e^{-r/L} r^2 dr \int_0^\pi \sin\theta d\theta \; r\cos(k r \cos\theta)={3\pi \sigma \abs{\nabla \overline c}^2\over k^4}.
\end{eqnarray}
The remaining triple integral in  Eq.(\ref{C10}), aside from a factor of ${-3\over 16}\sigma\abs{\nabla\overline c}^2$, can be written as 
\begin{eqnarray}\nonumber&& \sin^2\alpha \int_0^{2\pi} d\phi \int_0^\infty e^{-r/L} r^3 dr \int_0^\pi \sin\theta \cos^2\theta\cos(kr\cos\theta) d\theta\\\nonumber
&&+ \cos^2\alpha\int_0^{2\pi} d\phi \int_0^\infty e^{-r/L} r^3 dr \int_0^\pi \sin^3\theta \cos^2\phi \cos(kr\cos\theta) d\theta\\\nonumber
&&+\int_0^{2\pi} d\phi \int_0^\infty e^{-r/L} r^3 dr\int_0^\pi d\theta \sin^2\phi  \sin^3\theta \cos(kr\cos\theta)\\\label{C20}
&&+ \sin2\alpha\int_0^{2\pi} d\phi \int_0^\infty e^{-r/L} r^3 dr  \int_0^\pi \sin^2\theta \cos\theta\cos(kr\cos\theta) d\theta.
\end{eqnarray}
In the last line, the integral over $\theta$ vanishes because the integrand is odd under the reflection $\theta\to \pi-\theta$. As for the remaining integrals in Eq.(\ref{C20}), those over $\phi$ are elementary and those over $\theta$ can be readily evaluated using the following results:
\begin{eqnarray}\nonumber
&&\int_0^\pi \sin\theta \cos^2\theta\cos(kr\cos\theta) d\theta\\\label{first}
&&={2\over k^3 r^3}\Big[ (k^2 r^2-2)\sin(kr)+2kr\cos(kr)\Big],
\end{eqnarray}
which can easily be obtained by a change of variable, e.g., as $t=\cos\theta$, and then integrating by parts,
and also 
\begin{eqnarray}\nonumber
&&\int_0^\pi \sin^3\theta \cos(kr\cos\theta) d\theta={4\over k^3 r^3}\Big[ \sin(kr)-kr\cos(kr)\Big],
\end{eqnarray}
which can be checked by writing $\sin^3\theta=\sin\theta(1-\cos^2\theta)$, then using Eq.(\ref{first}) and the trivial result $\int \sin\theta\cos(kr \cos\theta))d\theta=2\sin(kr)/kr$. Substituting these results back into Eq.(\ref{C20}), we see that the remaining integrals over $r$ are in fact Laplace transforms of the functions $(kr)^2 \sin(kr)$, $\sin(kr)$ and $kr \cos(kr)$ which can be evaluated using standard tables of integrals e.g., \cite{Erdelyi1954}. In the limit $L\to+\infty$, we find $\int_0^\infty e^{-r/L} (kr)^2 \sin(kr) dr\to -2/k$ using \cite{Erdelyi1954} formula 4.7 (14); $\int_0^\infty e^{-r/L} \sin(kr) dr\to 1/k$ using \cite{Erdelyi1954} formula 4.7 (1) and finally $\int_0^\infty e^{-r/L} (kr) \cos(kr) dr\to -1/k$ using \cite{Erdelyi1954} formula 4.7 (57). Putting all this together, and using Eq.(\ref{1st}), we find
\begin{eqnarray}\nonumber
S(k)&=&{3\pi\sigma\over k^4}\abs{\nabla \overline c}^2-{3\over 16}\sigma\abs{\nabla\overline c}^2\Big[{8\pi\over k^4}(1+\cos^2\alpha)-{24\pi\over k^4}\sin^2\alpha \Big].
\end{eqnarray}
 Simplifying the above expression, using $\cos^2\alpha=k_\parallel^2/k^2$ and $\sin^2\alpha=k_\perp^2/k^2$, and plugging it into Eq.(\ref{C10}) leads to Eq.(\ref{structure2}), which is the desired result.

\end{appendices}


\bibliography{Decay}

\end{document}